\documentstyle[preprint,tighten,aps,floats]{revtex}

\def\ub{{\overline{u}}}
\def\vb{{\overline{v}}}

\def\be{\begin{equation}}
\def\ee{\end{equation}}
\def\ben{\begin{equation*}}
\def\een{\end{equation*}}
\def\bea{\begin{eqnarray}}
\def\eea{\end{eqnarray}}
\def\e{\epsilon}
\def\b{\beta}
\def\t{\tau}

\begin{document}

\draft

\title{Five-loop $\epsilon$ expansion for $U(n)\times U(m)$ models: 
finite-temperature phase transition in light QCD}

\author{
Pasquale Calabrese${}^{1}$ and  
Pietro Parruccini${}^2$}  
\address{$^1$Theoretical Physics, University of Oxford, 1 Keble Road, 
Oxford OX1 3NP, United Kingdom.}
\address{$^{2}$ Dipartimento di Chimica Applicata, Universit\`a di Bologna, 
Via Saragozza 8, I-40136 Bologna, Italy.
}

\date{\today}

\maketitle

\begin{abstract}

We consider the  $U(n)\times U(m)$ symmetric $\Phi^4$ Lagrangian to describe
the finite-temperature phase transition in QCD in the limit of vanishing 
quark masses with $n=m=N_f$ flavors and unbroken anomaly at $T_c$.  
We compute the Renormalization Group functions to five-loop order in
Minimal Subtraction scheme.
Such higher order functions allow to describe accurately the three-dimensional
fixed-point structure in the plane $(n,m)$, and to reconstruct the line 
$n^+(m,d)$ which limits the region of second-order phase transitions
by an expansion in $\epsilon=4-d$.
We always find $n^+(m,3)>m$, thus no three-dimensional
stable fixed point exists for $n=m$ and 
the finite temperature transition in light QCD should be first-order.
This result is confirmed by the pseudo-$\e$ analysis of massive 
six-loop three dimensional series.

\end{abstract}


\section{Introduction}

The phase diagram of QCD is characterized by a low temperature hadronic
phase with broken chiral symmetry and an high temperature phase with
deconfined quarks and gluons, in which chiral symmetry is restored.
The nature of the transition between these two phases depends on the QCD
parameters, as the number of flavors and quark masses. In the limit of
zero quark masses such phase transition  is essentially related to the
restoring of chiral symmetry (see e.g. the reviews \cite{rev}).

The QCD Lagrangian with $N_f$ massless quarks is classically invariant 
under the global flavor symmetry 
$U(1)_A \times SU(N_f) \times SU(N_f)$ \cite{pw-84}.
The axial $U(1)_A$ symmetry may be broken by the anomaly at the quantum level,
reducing the relevant symmetry to
$SU(N_f)\times SU(N_f)\times {Z}(N_f)_A$ \cite{pw-84}.
At $T=0$ the symmetry is spontaneously broken to $SU(N_f)_V$ with a nonzero 
quark condensate. With increasing $T$, a phase transition characterized by 
the restoring of the chiral symmetry is expected at $T_c$.
To parameterize this phase transition a complex $N_f$-by-$N_f$ matrix 
$\Phi_{ij}$ is introduced as an order parameter.
The most general renormalizable three-dimensional $U(N_f)\times U(N_f)$ 
symmetric Lagrangian is 
\cite{pw-84,w-92}
\be
{\cal L}_{U(N_f)}={\rm Tr} (\partial_\mu \Phi^\dagger) (\partial_\mu \Phi)
+r {\rm Tr} \Phi^\dagger \Phi 
+ {u_0\over 4} \left( {\rm Tr} \Phi^\dagger \Phi \right)^2
+ {v_0\over 4} {\rm Tr} \left( \Phi^\dagger \Phi \right)^2 \,,
\label{LGWL}
\ee
which describes the QCD symmetry breaking pattern only if 
$v_0>0$ \cite{bpv-03}.

If the anomaly is broken at $T_c$, the effective Lagrangian 
is \cite{pw-84,w-92} 
\be
{\cal L}_{SU(N_f)}={\cal L}_{U(N_f)}+ w_0(\det \Phi+\det\Phi^{\dagger}).
\label{AT}
\ee
The effect of non-vanishing quark masses can be accounted for by adding a 
linear term $m_{ij} \Phi_{ij}$ in the Lagrangian \cite{rev,pw-84,w-92}, 
that acts as a magnetic field in a spin model.

The mean-field analysis of the $U(N_f)\times U(N_f)$ Lagrangian 
Eq. (\ref{LGWL})
predicts a second-order phase transition everywhere the stability conditions
$v_0\geq 0$ and $N_f u_0+v_0\geq 0$ are satisfied. However, according to
Renormalization Group (RG) theory the critical behavior at a
continuous phase transition is described by the stable fixed point (FP)
of the theory \cite{zj}. 
The absence of a stable FP indicates that the transition
cannot be continuous, even though mean-field suggests it. 
Therefore  the transition is expected to be first-order (see e.g.
\cite{rev-01}).

The $U(N_f)\times U(N_f)$ Lagrangian was studied at one-loop in 
$\e=4-d$ expansion \cite{pw-84,ps-81,pat-81} and at six-loop in the massive 
zero-momentum renormalization scheme directly in $d=3$ \cite{bpv-03}. 
No stable FP was found for all values of $N_f\geq2$, concluding for a 
first-order phase transition.
Anyway both the used approximations have intrinsic limits. 
The one-loop $\e$ expansion, as discussed in Ref. \cite{bpv-03},
provides useful qualitative indications for the description of the RG 
flow, but it fails in describing quantitatively 
the right three-dimensional behavior.
The  major drawback of fixed dimension expansion is that the 
numerical resummation techniques necessary to 
extract quantitative informations allow to explore a large, but 
limited, region in the space of coupling constants~(e.g. in 
Ref. \cite{bpv-03} the resummation results to be effective only in the 
region $-2\leq \ub,\vb \leq 4$, where $\ub,\vb$ are the couplings
used in Ref \cite{bpv-03}). 
One cannot exclude a priori that a FP may be outside the accessible region of 
effectiveness of resummation in $d=3$.
These problems are absent in $\e$ expansion since no resummation is 
needed to find the FP's, being series in $\e$.

For these reasons, we extend the $\e$ expansion series to 
five loops. We consider the $U(n)\times U(m)$ generalization of Lagrangian
(\ref{LGWL}), where $\Phi$ is a $n$-by-$m$ complex matrix.
To understand why we decide to study this more general model 
let us consider the already known one-loop $\beta$ functions \cite{ps-81}:
\bea
\beta_u(u,v)&=&-\e\, u + (n m + 4) u^2 + {2}(n+m) u v + 3v^2\,,\quad
\beta_v(u,v)=-\e\, v + 6 u v +{(n + m)} v^2\,.
\label{1loop}
\eea
For $n=m$ a couple of FP's with non-vanishing and negative $v$ exists only
for $n<\sqrt3+O(\e)$ \cite{pw-84}, suggesting a first-order
phase transition for those systems, as light QCD, having $v_0>0$.
Anyway, if one considers the model with $n\neq m$ 
a more complicated structure of FP's emerges.
Two FP's, the Gaussian one $(u^*=v^*=0)$ and the $O(2 n m)$  one $(v^*=0)$, 
always exist. For $n \geq n^+(m,d)$ and $n\leq n^-(m,d)$ other two FP's 
appear which we call $U^+$ and $U^-$, whose coordinates at one-loop read
\be
u^*_{\pm}=\case{A_{mn}\pm (m+n) R_{mn}^{1/2}}{2 D_{mn}} \e,
\qquad
v^*_{\pm}=\case{B_{mn}\mp 3 R_{mn}^{1/2}}{D_{mn}}\e,
\ee
with
\bea
B_{mn}&=&n m^2-5n+m n^2-5m\,, \qquad 
A_{mn}=36-m^2-2 mn- n^2\,, \\
R_{mn}&=&24 + m^2 - 10 m n + n^2\,,\quad 
D_{mn}=108 - 8 m^2 - 16 m n + m^3 n - 8 n^2 + 2 m^2 n^2+ m n^3\,.\nonumber
\eea
The $v^*$ coordinates of $U^\pm$ for
$n>n^+$ are positive, as it should be to provide the right symmetry
breaking of QCD. 
Requiring $R_{mn}>0$, 
we have $n^\pm=5m \pm2\sqrt{6} \sqrt{m^2-1}+O(\e)$\footnote{Note that it is 
possible to study the generalized $U(n)\times U(m)$ 
model at fixed $m$ in a $1/n$ expansion, since it has a FP for $n=\infty$,
contrarily to the $U(n)\times U(n)$.}.

The stability properties of these FP's at fixed $m$ (for the physically 
relevant case $m \geq 1$) are characterized by the following four different 
regimes (note the analogy with the $O(n)\times O(m)$ 
model \cite{Kawamura-98}):
\begin{itemize}
\item[1)] For $n>n^+(m,d)$, there are four FP's, and $U^+$
is the only stable. Both $U^\pm$ have $v^*>0$. 
\item[2)] For $n^-(m,d) < n < n^+(m,d)$, only the Gaussian and the Heisenberg
O($2n m$)-symmetric FP's are present, and none of them is stable.
Thus the transition is expected to be first-order.
\item[3)] For $n_H(m,d) < n < n^-(m,d)$, there are again four FP's,
and $U^+$ is the stable one.
However at small $\e$ for $m<5$, it has $v^*<0$. For $v_0>0$ a first-order 
transition is  expected. For this reason we will never consider 
the value of $n^-(m,d)$.
\item[4)] For $n < n_H(m,d)$, \footnote{The value of $n_H(m)$ may be inferred 
from Refs. \cite{cpv-02m,cpv-03r}, where it was shown, on the basis that at a 
global $O(N)$ FP all the spin four operators have the same scaling dimension,
that the $O(N)$ FP is stable only if
$N<N_c$, with $N_c\sim2.9$ \cite{rev-01,cpv-03r,nc}.
Thus the $O(2nm)$ FP is stable for $n\alt 1.45/m$ and it is not 
expected to play any role at the QCD phase transition.}
there are again four FP's,
and the Heisenberg O$(2m n)$-symmetric one is stable. 
\end{itemize}

Now it is possible (and one has to check!) that the actual value of
$n^+(m,3)$ is lower than $m$, providing a stable FP and consequently a new 
universality class for $U(N_f)\times U(N_f)$ symmetric models.
To give a definitive answer to this question, high order calculations are 
required, since low order ones lead to erroneous conclusions, as we shall see.
However we anticipate that we do not find any stable FP, supporting
the results of Ref. \cite{bpv-03}.

The paper is organized as follows. 
The $U(n)\times U(m)$ model is analyzed at five-loop in $\e$ expansion in 
Sec. \ref{sec3}. 
In Sec. \ref{sec3.5}, the model is reanalyzed with pseudo-$\e$ expansion 
at six-loop order in massive zero-momentum renormalization scheme.
Sec. \ref{sec5} summarizes our main results. 
In the appendix we briefly discuss the effect of the anomaly.

\section{Five-loop $\epsilon$-expansion of $U(n)\times U(m)$ model.}  
\label{sec3}

We extend the one-loop $\epsilon$ expansion of Refs.
\cite{ps-81} for the RG functions of the $U(n)\times U(m)$ 
symmetric theory to five-loop.
For this purpose, we consider the minimal subtraction ($\overline{\rm MS}$) 
renormalization scheme \cite{zj} for the massless theory.
We compute the divergent part of
the irreducible two-point functions of the field $\Phi$,
of the two-point correlation functions with insertions of the 
quadratic operators  $\Phi^2$, and 
of the two independent four-point correlation functions.
The diagrams contributing to this calculation are 162 for the four-point 
functions and 26 for the two-point one.
We handle them with a symbolic manipulation program, which  
generates the diagrams and computes  the symmetry and group factors of 
each of them. We use the results of  Ref.~\cite{KS-01}, 
where the primitive divergent parts of all integrals appearing in our 
computation are reported.
We determine the renormalization constant $Z_{\Phi}$ 
associated with the fields $\Phi$, the renormalization constant  $Z_t$ of the 
quadratic operator $\Phi^2$, and the renormalized quartic couplings $u,v$.
The functions $\beta_u$, $\beta_v$, $\eta_\phi$ and  $\eta_t$ are 
determined using the relations
\be
\beta_u (u,v) = \mu \left. {\partial u \over \partial \mu} \right|_{u_0,v_0},
\qquad\qquad
\beta_v (u,v) = \mu \left. {\partial v \over \partial \mu} \right|_{u_0,v_0},
\ee
\begin{eqnarray}
\eta_\phi (u,v)&=& 
  \left. {\partial \log Z_\Phi \over \partial \log \mu} \right|_{u_0,v_0},
\qquad
\eta_t (u,v)= \left. {\partial \log Z_t \over \partial \log \mu} \right|_{u_0,v_0}.
\end{eqnarray}

The zeroes $(u^*,v^*)$ of the $\beta$ functions provide the FP's
of the theory. In the framework of the $\epsilon$ expansion, they are 
obtained as perturbative expansions 
in $\epsilon$ and then are  inserted in the RG functions to determine
the $\epsilon$ expansion of the critical exponents.
\be
\eta=\eta_\phi (u^*,v^*), \quad
\nu =(2 - \eta_\phi (u^*,v^*) - \eta_t(u^*,v^*))^{-1}.
\ee

\subsection{RG functions}

The five-loop expansions of the $\beta$ functions are given by
\bea
\beta_u&=&-\e\, u + (n m + 4) u^2 + 2(n+m) u v +3v^2
-\case{3}{2}(7 + 3 m n ) u^3 - 11 (m+ n) u^2 v \nonumber\\
&&- \case{ 41 + 5 m n }{2} u v^2  - 3( m + n) v^3
+\left(\case{ 740 + 461 m n + 33 m^2 n^2 }{16} +\zeta(3)(33+15 mn)\right) u^4 \nonumber\\
&&+\left(\case{659 n + 79 m^2 n +  659 m+ 79 m n^2 }{8}+48 \zeta(3)(m+n)\right)u^3 v\nonumber\\
&&+\left(\case{2619 + 1210 m n + 230 n^2 + 230 m^2+ 3 n^2 m^2}{16}+18\zeta(3)(7+m n)\right) u^2 v^2 
\nonumber\\
&&+ \left(\case{15}{4}(20 n + m^2 n + 20 m + n^2 m)+36 \zeta(3)(m+n)\right) u v^3 \nonumber\\
&&
+ \left(\case{ 425 + 20 m^2 + 153 m n + 20 n^2}{16}+6\zeta(3)(4+m n)\right) v^4
+\b_5^u+\b_6^u\eea
\bea
\beta_v(u,v)&=&-\e\, v + 6 u v +(n + m) v^2 
-\case{41+5m n}{2}u^2 v-11(m+n)uv^2-3\case{5+m n}{2}v^3\nonumber\\
&&+v\left[\left(\case{821+184 mn -13 m^2 n^2}{8}+12\zeta(3)(7+m n)\right) u^3\right.\nonumber\\
&&\left.
+\left(\case{1591\,n - 35\,m^2\,n + 1591 m- 35 m n^2}{16} +72\zeta(3)(m+n)\right)u^2 v \right.\nonumber\\
&&\left.
+\left(\case{211 + 9\,m^2 + 73 m n + 9 n^2}{2}+24\zeta(3)(4+m n)\right) u v^2 \right.\nonumber\\
&&\left.
+\left( \case{295 n + 13 m^2 n +  295 m+ 13 m n^2}{16}+9\zeta(3)(m+n)\right) v^3 \right]
\,+\b_5^v+\b_6^v.
\eea
The coefficients $\b_5^u,\b_6^v,\b_5^u,\b_6^u$ are very long and not really 
illuminating. We do not report them here, but they are available
on request to the authors.
The same is true for the RG functions $\eta_\phi$ and $\eta_t$ to five-loop,
that we calculated but never used, since we did not find evidence 
for a FP in the space of parameters of interest.

We have checked that for $v=0$ the series reduce to the existing 
$O(\epsilon^5)$ ones for the O($2nm$)-symmetric theory \cite{ON}.
For $m=1$ and any $n$ (or viceversa, given the symmetry under the exchange 
$n\leftrightarrow m$) the theory is equivalent to an $O(2n)$ in the 
coupling $u+v$, so the series satisfy the relation
$\beta_u(z+y, z-y;n,m=1) + \beta_v(z+y, z-y;n,m=1) = \beta_{O(2n)}(z),$ 
where $\beta_{O(2n)}(z)$ is the $\beta$-function of 
the O($2n$) model \cite{ON}.

\subsection{Estimates of $n^+(m,3)$}

From the above reported series, the $\epsilon$ expansion of 
$n^\pm(m)$ may be calculated to $O(\epsilon^5)$. 
$n^\pm(m)$ may be expanded as
\be
n^\pm(m) = n_0^\pm(m) + n_1^\pm(m)\, \epsilon + n_2^\pm(m)\, \epsilon^2  
 +n_3^{\pm}(m)\, \e^3+n_4^{\pm}(m)\, \e^4   +      O(\epsilon^5),
\ee
and the coefficients $n_i^\pm(m)$ are obtained by requiring 
\be
\beta_u(u^*,v^*;n^\pm) = 0, \qquad 
\beta_v(u^*,v^*;n^\pm) = 0, 
\quad {\rm and}\qquad 
{\rm det}\, \left|{\partial(\beta_u,\beta_v)\over \partial(u,v)} \right| 
 (u^*,v^*;n^\pm) = 0.
\ee
For generic values of $m$, the expression of $n^\pm(m,4-\e)$ is too
cumbersome in order  to be reported here. 
We only report the numerical expansion of $n^+$ at fixed $m=2,\,3,4$, i.e.
\bea
n^+(2,4-\e)&=&18.4853-19.8995 \e+2.9260 \e^2+4.6195 \e^3-0.7182 \e^4+O(\e^5)\,,\nonumber\\
n^+(3,4-\e)&=&28.8564-30.0833 \e+6.5566 \e^2+3.4056 \e^3-0.7958 \e^4+O(\e^5)\,,\nonumber\\
n^+(4,4-\e)&=&38.9737-40.2386 \e+9.6089 \e^2+3.0505 \e^3-0.6156 \e^4+O(\e^5)\,,
\label{n+exp}
\eea
In order to give an estimate of $n^+(m,3)$ such series should be
evaluated at $\e$=1. 
A linear extrapolation of the two-loop contribution leads to the wrong 
conclusion that $n^+(m,3)<m$, i.e. that the transition is continuous.
This is the anticipated statement that high-loop computation are needed to 
have a conclusive result. 
Anyway, a direct sum of the five-loop series is not effective, since they 
are expected to be divergent.
The high irregular behavior with the number of the loops makes also a
Borel-like resummation non effective as well (in fact Pad\'e-Borel 
resummation leads to unstable results). 
Thus we try to extract from Eqs. (\ref{n+exp}) better behaved series by means 
of algebraic manipulations.

This may be done considering (as in Ref. \cite{cp-03})
\bea
1/n^+(2,4-\e)&=&0.0541+0.0582\e+0.0541\e^2+0.0355\e^3 + 0.0172\e^4+O(\e^5)\,,\nonumber\\
1/n^+(3,4-\e)&=&0.0347+0.0361\e+0.0298\e^2+0.0188\e^3 + 0.0095\e^4+O(\e^5)\,,\nonumber\\
1/n^+(4,4-\e)&=&0.0257+0.0265\e+0.0210\e^2+0.0132\e^3 + 0.0067\e^4+O(\e^5)\,,
\eea
whose coefficients decrease rapidly. Setting $\e=1$ we obtain the
results reported in Table \ref{Tabn+}.

\begin{table}[tbp]
\caption{Estimates of $n^+$ for several $m$ with varying the number of loops.}
\label{Tabn+}
\begin{center}
\begin{tabular}{c|lll|lll|l}
\multicolumn{1}{c|}{m}& 
\multicolumn{3}{c|}{$1/n^+$}&
\multicolumn{3}{c|}{$a$}&
\multicolumn{1}{c}{final}\\
& 3-loop& 4-loop & 5-loop & 3-loop& 4-loop & 5-loop &\\
\hline
2& 6.01 & 4.95 & 4.56 & 4.98 & 4.55 & 4.44 & 4.5(5)\\
3& 9.94 & 8.38 & 7.76 & 8.07 & 7.53 & 7.40 & 7.6(8)\\
4& 13.7 & 11.6  & 10.7 & 11.0 & 10.3 & 10.2 & 10.5(1.1)\\
\end{tabular}\end{center}\end{table}

Another method, firstly employed for $O(m)\times O(n)$ models in 
Ref. \cite{prv-01n}, use the knowledge of $n^+(m,2)$ to constrain the 
analysis at $\e=2 $, under the (strong) assumption that 
${n^+}(m,d)$ is  sufficiently smooth in $d$ at fixed $m$.
$n^+(m,2)$ may be conjectured further assuming that the two-dimensional LGW
stable FP is equivalent to that of the NL$\sigma$ model for all
$n\ge 1$ except $n=1$ \cite{zj}. Since  the NL$\sigma$ model is
asymptotically free, we conclude that $n^+(m,2)=1$.
The knowledge of ${n}^+(m,2)$ may be exploited in order to 
obtain some informations on $n^+(m,3)$, rewriting the perturbative series 
for $n^+(m,4-\e)$ in the following form
\be
n^+(m,4-\e)=1 + (2 - \e)\,a(m,\e),
\ee
where
\bea
a(2,\e)&=& 8.743 - 5.579\e - 1.326\e^2 + 1.646\e^3 + 0.464\e^4 +O(\e^5)\,,\nonumber\\
a(3,\e)&=&13.928 - 8.078\e - 0.760\e^2 + 1.323\e^3 + 0.263\e^4 +O(\e^5)\,,\nonumber\\ 
a(4,\e)&=&18.987 - 10.626\e - 0.508\e^2 + 1.271\e^3 + 0.328\e^4 +O(\e^5)\,,
\eea
whose terms are better behaved than the original series, but not decreasing.
We consider the  series $a(2,\e)^{-1}$ obtaining a more ``convergent'' 
expression, which may be estimated simply by setting $\e=1$. 
The results of this constrained analysis are reported in Table \ref{Tabn+}.
Note that, although the several not completely justified assumptions we
made, the final obtained series are highly stable with changing the number
of the loops. Obviously this is not an evidence favoring the validity of
the assumptions, but it is a very convincing argument to ensure the
goodness of our estimates.

As final estimates we quote  an average of the five-loop results
(that are quite close) and as error bar the maximum difference with the
fourth order ones. 
For all the considered value of $m$ we have $n^+(m,3)>m$, thus the
transition for $U(n)\times U(n)$ models is expected to be first-order.
We also check that $n^+(m,3)>m$ for higher values of $m$.

Since the new couple of FP's does not exist for finite temperature 
transition of light QCD, we do not report their expansion in terms of $\e$ 
and the exponents characterizing the critical behavior for $n>n^+$. 
Anyway they may be obtained from the series we reported and from those that 
are available on request.

\section{Pseudo-$\e$ expansion}
\label{sec3.5}

In this section we analyze the six-loop zero-momentum massive 
three-dimensional series with the so-called  pseudo-$\e$ expansion 
method \cite{lz-80}, since it provided the best results 
for the marginal spin dimensionality in spin models (see e.g. Refs. 
\cite{cp-03,rev-01} and references therein). 
The $\b$ functions for generic $n$ and $m$ were calculated at six-loop
in Ref. \cite{bpv-03} (but they were not reported there).

The idea behind the pseudo-$\e$ expansion is very simple \cite{lz-80}: 
one has only to multiply the 
linear terms of the two $\beta$ functions by a parameter $\t$, find the 
the common zeros of the $\beta$'s as series in $\t$ and analyze 
the results as in the $\e$ expansion.
The critical exponents are obtained as series 
in $\t$ inserting the FP's expansions in the appropriate RG functions. 
Note that, differently from the $\e$ expansion, only the value 
at $\t=1$ makes sense. 
The reason for which it works well is twofold: first in the three 
dimensional approach at least one order more in the loop expansion is known,
second, and more important, the RG functions are better behaved in the 
massive approach \cite{zj,lz-80}.

\begin{table}[!t]
\begin{center}
\caption{\small Pad\'e table for $n^+(2,3)$ in pseudo-$\e$ expansion. 
The location of the positive real pole closest to the origin is reported in 
brackets.}
\begin{tabular}{l|cccccc}
& $N=0$ & $N=1$ &$N=2$ &$N=3$ &$N=4$ &$N=5$ \\
\hline
$M=0$&18.485 &5.219&4.769&4.596&4.542&4.527\\
$M=1$&10.762 &$4.753_{[29.5]}$ &$4.487_{[2.59]}$&$4.518_{[3.23]}$&
$4.522_{[3.72]}$&\\
$M=2$&8.190  &$4.579_{[7.43]}$ &$4.518_{[3.21]}$&4.523&&\\
$M=3$&6.921  &$4.533_{[4.99]}$ &$4.522_{[3.65]}$&&&\\
$M=4$&6.181  &$4.524_{[4.30]}$ &&&&\\
$M=5$&5.708&&&&&\\
\end{tabular}
\label{paden+}
\end{center}
\end{table}

Following the recipe explained in the previous section for the 
$\overline{\rm MS}$ scheme, we obtain 
\bea
n^+(2,3)&=&18.4853-13.2663 \t-0.4499\t^2-0.1735\t^3-0.0537\t^4-0.0144\t^5\,,\nonumber\\
n^+(3,3)&=&28.8564-20.0555 \t-0.3092\t^2-0.2609\t^3-0.1444\t^4-0.0968\t^5\,,\nonumber\\
n^+(4,3)&=&38.9737-26.8257 \t-0.2690\t^2-0.3582\t^3-0.2199\t^4-0.1526\t^5\,.
\eea
At least up to the known order, such expansions do not
behave as asymptotic with factorial growth of coefficients and 
alternating signs. So one may apply a simple Pad\'e resummation \cite{cp-03}.
The results of the $[N/M]$ Pad\'e approximants are displayed in Tab. 
\ref{paden+} for $m=2$ . 
Several approximants have poles on the positive real axis.
Anyway all these poles are ``far'' from $\t=1$, 
where the series must be evaluated. Thus one may expect 
the presence of a pole not to influence the result at $\t=1$. 
Anyway for the cases $m=3,4$ some Pad\'e have poles at $\t<2$,
that must be discarded in the average procedure.
We choose as final estimate the average the six-loop order Pad\'e
without poles at $\t<2$
(excluding those with $N=0$, giving unreliable results), and 
as error bar we take the maximum deviation from the average of four- and 
five-loop  Pad\'e (as in Ref. \cite{cp-03}).
Within this procedure we have $n^+(2,3) = 4.52(7)$, 
$n^+(3,3)= 7.98(25)$, $n^+(4,3)= 11.1(4)$.

The final results are in very good agreement with those of the previous 
section from a completely different approach. This is a clear evidence that
the estimates we made are robust.

\section{Conclusions}
\label{sec5}

In this paper we investigated the possibility of a second-order
phase transition in QCD in the limit of vanishing masses.
When the $U(1)_A$ symmetry is restored at $T_c$, the finite
temperature chiral transition, if continuous, may be described by the
Lagrangian ${\cal L}_{U(N_f)}$ Eq. (\ref{LGWL}). 
We pointed out that the correct extrapolation at $\e=1$ is obtainable
considering its $U(n)\times U(m)$ symmetric extension at fixed $m$. 
In fact the last model has a stable FP 
for $n>n^+(m,d)$ that is not accessible from the theory with $n=m$. 
The presence of a stable FP for light QCD with $N_f=m$ flavors requires 
$n^+(m,3)>m$. 
After showing that low order calculations are not conclusive,
we performed a five-loop expansion that allowed to conclude that no 
continuous transition is possible for three dimensional models with 
$U(n)\times U(n)$ symmetry for $n\geq 2$.
We corroborated this result with a direct three-dimensional analysis,
namely with the pseudo-$\e$ expansion at six-loop order.

In Ref. \cite{bpv-03} six-loop massive zero momentum series were analyzed 
directly in three dimensions allowing to exclude that a FP, without 
any counterpart in $\e$ expansion exists
(as it was claimed to happen for $O(n)\times O(2)$ 
models for low values of $n$ both for $v>0$ \cite{prv-01p} 
and $v<0$ \cite{dpv-03}).
We believe that this work, together with Ref. \cite{bpv-03}, put on a robust 
basis the prediction that the transition in $U(n)\times U(n)$ 
models is first-order.

In the appendix, following Ref. \cite{bpv-03}, we point out
that the anomaly may lead to a continuous transition only for $N_f=2$ and 
large values of $|w_0|$, instead for $N_f\geq 3$ it does not 
softens the first-order transition of the $U(N_f)\times U(N_f)$ model, for 
any value of $w_0$.

Finally it is worth mentioning that the $U(n)\times U(m)$ models could be 
relevant in the description of some quantum phase transitions, as it happens
for their $O(n)\times O(m)$ counterparts for Mott insulators \cite{sach}.
Being $n^+(m,3)>m$ for all $m\geq2$, we predict a first-order phase transition
for all those systems with $m\leq3$, which are interesting for the 
condensed matter point of view.

\section*{ACKNOWLEDGMENTS}

We thank Luigi del Debbio and Ettore Vicari for useful discussions.
We thank E. Vicari for providing us six-loop three dimensional RG functions
calculated and not reported in Ref. \cite{bpv-03}. 
PC acknowledges financial support from EPSRC Grant No. GR/R83712/01.

\appendix

\section{The effect of the anomaly}

If the $U(1)_A$ symmetry is broken at $T_c$ by the anomaly, one has to
consider the Lagrangian (\ref{AT}).
Since the effect of the anomaly is always (apart from the case $N_f=4$) well
described by general arguments, this appendix is very similar to part of 
Ref. \cite{bpv-03}. However, we report such arguments here in order to
make this paper self-consistent.

In the large-$N_c$ limit (where $N_c$ is the number of colors),
the effect of the anomaly tends to be suppressed, and the Lagrangian 
Eq. (\ref{LGWL}) is recovered in the limit $N_c\rightarrow\infty$. 
The effective $U(1)_A$-symmetry breaking
at finite temperature in real QCD has been investigated on the lattice.
The $U(1)_A$ symmetry appears not to be restored at $T_c$, but
the effective breaking of the axial $U(1)_A$ symmetry 
appears substantially reduced especially above $T_c$ 
(see, e.g., Refs.~\cite{U1}).
However, the Lagrangian (\ref{LGWL}) still describe a large part of the phase
diagram of the model with broken anomaly as we show closely following Ref. 
\cite{bpv-03}.

For $N_f=2$ the symmetry breaking pattern is equivalent to
$O(4)\rightarrow O(3)$ \cite{pw-84}. If the transition is continuous it
is in the three-dimensional $O(4)$ universality 
class \cite{pw-84,w-92,rw-93,ggp-94}, which has been accurately studied in 
the literature \cite{rev-01,O4}. 
Actually a continuous transition is expected only for large enough value of 
$|w_0|$ (see Ref. \cite{bpv-03}, in particular Appendix A, for the phase 
diagram of this model).
In particular the multicritical point is $U(2)\times U(2)$ symmetric.
The phase diagram realized in light QCD may be understood only
from the QCD Lagrangian and not from universality arguments. Lattice
simulations in two flavors QCD favor a continuous transition consistent
with the $O(4)$ universality class \cite{latticen2}.

For $N_f=3$ the determinant is cubic in $\Phi$, making the Lagrangian not
bounded. So the transition is expected to be first-order for all $w_0$. 
Lattice simulations of QCD confirm this expectation \cite{latticen3}.

For $N_f=4$ the determinant is quartic in $\Phi$, leading to the three
couplings effective $\Phi^4$ Lagrangian
\be
{\cal L}_{SU(4)}={\cal L}_{U(4)}+w_0 \e_{ijkl}\e_{abcd}\Phi_{ia} \Phi_{jb} \Phi_{kc} \Phi_{ld}\,, 
\ee
where $\epsilon_{ijkl}$ is the completely antisymmetric tensor 
($\epsilon_{1234}=1$). Such Lagrangian is not generalizable to an
$n$-by-$m$ matrix with $n$ or $m$ different from 4 and so we limit to 
consider the case $n=m=4$.
The one-loop $\beta$ functions we obtain are
\begin{eqnarray}
\beta_u (u,v,w)&=&-\e u+ 20 u^2 +{16} u v +{3} v^2 + {8}  w^2\,,\nonumber\\
\beta_v (u,v,w)&=&-\e v + {8}  v^2 +{6}  u  v   - {8}  w^2\,,\nonumber\\
\beta_w (u,v,w)&=&-\e w +{6}  u  w -{6}  v  w\,.
\end{eqnarray}
Such series, as in six-loop three-dimensional case \cite{bpv-03}, have no 
common zeros with non-vanishing coordinates. 
Anyway, differently from three dimensions, the $\beta_w$ vanishes
in a region of parameters, different from $w=0$ (namely the 
surface $\e-6( u- v)=0$). 
Higher loop corrections may not change the number of FP's,
since they are expansions in $\e$.

For $N_f\geq 5$ the added anomaly term is irrelevant since it generates
polynomials of degrees higher than four. Therefore for $N_f\geq 5$  the 
Lagrangians ${\cal L}_{SU(N_f)}$ and ${\cal L}_{U(N_f)}$ are equivalent 
at criticality.

\end{document}